\documentclass[aps,twocolumn,superscriptaddress]{revtex4}
\usepackage{graphicx}  % needed for figures
\usepackage{dcolumn}   % needed for some tables
\usepackage{bm}        % for math
\usepackage{amssymb}   % for math
\usepackage{hyperref}
\usepackage{tikz}
\usetikzlibrary{shapes,arrows}
\usepackage{amsmath}
\begin{document}

\title{Entropic force and bouncing behaviour in $\kappa$-Minkowski space-time}

\author{Vishnu Rajagopal}
\email{vishnu@hunnu.edu.cn}
\affiliation{Department of Physics and Synergetic Innovation\\
Center for Quantum Effects and Applications,\\
Hunan Normal University, Changsha, Hunan 410081, China}
\affiliation{Institute of Interdisciplinary Studies,\\
Hunan Normal University, Changsha, Hunan 410081, China}
\author{Puxun Wu}
\email{pxwu@hunnu.edu.cn}
\affiliation{Department of Physics and Synergetic Innovation\\
Center for Quantum Effects and Applications,\\
Hunan Normal University, Changsha, Hunan 410081, China}
\affiliation{Institute of Interdisciplinary Studies,\\
Hunan Normal University, Changsha, Hunan 410081, China}

\begin{abstract}
We generalise the entropic force description of gravity into $\kappa$-Minkowski space-time and derive the $\kappa$-deformed corrections to the Newton's gravitational force. Using this we show the appearance of logarithmic correction as the first order $\kappa$-deformed correction term to Bekenstein-Hawking entropy. Further we also derive the $\kappa$-deformed Friedmann equations and study the evolution of scale factor in $\kappa$-deformed space-time. Finally we show that the $\kappa$-deformation parameter avoids the initial singularity in early universe by providing a bounce behaviour for the case of spatially flat and closed universe.

\end{abstract}

\maketitle

\section{Introduction}

The classical description of gravity is known to break down at the singularities and hence the resolution of singularity is one of the major issues manifested in the general theory of relativity. A consistent quantum theory of gravity is capable of solving the singularity problems in general relativity. The quantum gravity theories feature the existence of a fundamental minimal length scale \cite{ml} and thus various approaches have been initiated in the recent times to capture the quantum gravity effects by incorporating this minimal length into the prevailing theories. The loop quantum gravity (LQG) is one such approach which utilises the quantum geometry framework to provides a plausible resolution of the cosmological singularity by replacing the big bang with a big bounce scenario \cite{bojowald,abhay}. However the LQG and other prominent quantum gravity descriptions such as string theory, etc., are still insufficient to answer various issues of the quantum gravity. The quantum gravity effects have been shown to be significant around the Planck scale and most of these quantum gravity approaches have predicted the existence of fundamental length scale related to the Planck scale. Thus one could possibly observe some signatures of quantum gravity effects by incorporating this minimal length scale into the theories.

The non-commutative (NC) geometry framework, which naturally contains the minimal length scale, is another way to look for the quantum gravity effects \cite{connes}. The quantum space-time structure, with a non-commutative algebra, has been shown to model the space-time uncertainties (carrying quantum gravity effects), obtained by combining the Einstein's theory of gravity and Heisenberg's uncertainty principle \cite{doplicher}. Thus replacing the standard space-time manifold with a NC space-time structure is an intuitive attempt to capture the quantum gravity effects. Subsequently different types of NC space-time structures have been constructed thoroughly and various consistent physical models have been studied in it.

The $\kappa$-Minkowski space-time is one of the major NC space-time which has been extensively studied in the recent times whose underlying space-time algebra is \cite{kappa-review}
\begin{equation} \label{lie}
 [\hat{x}_i, \hat{x}_j]=0,~~[\hat{x}_0,\hat{x}_i]=i\kappa^{-1}\hat{x}_i.
\end{equation}
In general the symmetry algebra of NC space-times are defined by the deformed Poincare algebra. For the $\kappa$-Minkowski space-time it is described by $\kappa$-Poincare algebra \cite{kpa} and this has been shown to be related with the Doubly Special Relativity (DSR) which treats the length scale $\alpha~(\equiv 1/\kappa)$ as a frame independent length scale \cite{dsr}. However it has been shown in \cite{meljanac} that symmetry algebra can also be defined using an undeformed $\kappa$-Poincare algebra such that standard Poincare algebra structure is retained but with a deformed co-algebra sector. Also this has been studied by employing the realisation approach where one realises the NC space-time coordinates as a function of commutative coordinate and its conjugate. Moreover it has been shown that there exist a one-to-one correspondence between the realisation choices and the ordering prescriptions \cite{kumar}. This approach facilitates us to study the physical problems with ease and therefore will use this approach in this work. 

Although various attempts to construct Einstein's gravitational theory in $\kappa$-deformed space-time have been made in \cite{kappa-gravity,kappa-gravity1}, the studies to construct a complete theory of $\kappa$-deformed general relativity, invariant under the deformed diffeomorphism symmetry, using the notion of $\kappa$-deformed differential calculus is still in progress. However the $\kappa$-Newton's gravity has been studied using different methods. In \cite{ehk}, the $\kappa$-deformed Newton's potential has been derived by replacing the commutative radial coordinate with its NC counterpart using a specific realisation choice compatible with the undeformed $\kappa$-Poincare algebra. The expression for the $\kappa$-deformed Newton's gravitational force has been obtained in \cite{ehk1} by taking the appropriate Newtonian limit of the geodesic equation in $\kappa$-Minkowski space-time.  

Different works have been carried out to understand the effects of non-commutativity in the very early universe. The effects of the $\kappa$-deformation on the evolution of early universe has been studied in \cite{suman} using the $\kappa$-deformed Friedmann equations, obtained by generalizing the conservation of energy and first law of thermodynamics to $\kappa$-Minkowski space-time. By using the modified Friedmann equations, obtained from the deformed Heisenberg algebra, the Snyder NC space-time has been shown to have non-singular bouncing universe \cite{snyder-bounce}. In \cite{moyal-bounce}, the cosmic evolution has been analysed, using the deformed Friedmann equation derived from the Lagrangian for NC teleparallel gravity after applying Seiberg-Witten map, and the bounce behaviour has been shown to appear in the Moyal NC space-time. 

Certain modified Friedmann equations containing the Planck scale corrections have also been shown to avoid the initial singularity of early universe. A cyclic non-singular universe has been shown to emerge from the generalised uncertainty principle (GUP) corrected Friedmann equation, derived by applying first law of thermodynamics and the GUP corrected entropy-area relation at apparent horizon \cite{gup-bounce}. The replacement of big-bang with a regular bounce has been suggested in \cite{alonso} while studying the evolution of the early universe due to the low-energy quantum gravity effects. Similarly the Planck scale corrected Friedmann equations obtained from the modified dispersion relation \cite{yi-ling,mdr-bounce} as well as from the rainbow gravity formalism \cite{rainbow-bounce} avoids the big bang singularity using the bouncing phenomenon. The bouncing universe has also been shown to appear in \cite{jusufi}, using the modified Friedmann equations, derived from the entropic force using a zero-point length scale corrected Newton's potential. These results are compelling enough to look for the possibility of a bounce scenario in $\kappa$-deformed space-time.

In this work we adopt the method discussed in \cite{verlinde} to derive the $\kappa$-Newton's laws of gravity. The Newton's gravity force in \cite{verlinde} has been considered to be an entropic force caused due to the changes in the entropy associated with the positions of material bodies, where the gravitating source has a holographic description in an emergent space scenario. We generalise this concept into $\kappa$-Minkowski space-time by considering the holographic screen to have a $\kappa$-corrected equilibrium temperature. It is worth noting that the expression for GUP-corrected Unruh temperature has been considered in \cite{gup-ef}, for studying the entropic gravity in GUP. Further by using this entropic force prescription we show that logarithmic correction term appears as the first order $\kappa$-deformed correction of Bekenstein-Hawking entropy. Additionally this entropic force method has also been utilised to obtain the standard Friedmann equations \cite{e-f,sheykhi}. We also utilise the similar procedure and derive the $\kappa$-deformed Friedman equations. By studying the evolution of early universe using the $\kappa$-corrected scale factor, obtained from the deformed Friedman equation, we finds that the $\kappa$-deformation parameter removes the initial singularity by providing a bouncing picture in flat as well as in closed universes.   

This paper is organised in the following manner. In Sec. II, we derive the $\kappa$-deformed corrections to Newton's potential using the entropic force method. Further we also calculate the $\kappa$-deformed corrections to Bekenstein-Hawking entropy. In Sec. III, we derive the $\kappa$-deformed Friedmann equations using the expression for the $\kappa$-deformed Newton's gravitational force. In Subsec. A, we study the evolution of early universe in $\kappa$-deformed space-time using the expression for $\kappa$-corrected scale factor, obtained by solving the modified Friedman equation.  In Subsec. B, we study the bouncing scenario in $\kappa$-deformed space-time. In Sec. IV, we discuss our results and provide the concluding remarks. Here we choose to work with $\hslash=c=k_B=1$.

%%%%%%%%%%%%%%%%%%%%%%%%%%%%%%%%%%%%%%%%%%%%%%%%%%%%%%%%%%%%%%%%%%%%%%%%%%%%%%%%%

\section{$\kappa$-deformed entropic force and Newton's gravity}

Here we study the Newtons' law of gravitation using the notion of entropic force in $\kappa$-Minkowski space-time. We explicitly calculate the $\kappa$-deformed corrections to the entropic force by utilising the expression for the $\kappa$-corrected Unruh temperature which we have derived in our earlier work \cite{vr1}. Using this we then obtain the $\kappa$-deformed corrections to Newton's potential. Further we also derive the logarithmic corrections to the Bekenstein-Hawking entropy just by considering the first order corrections associated with the $\kappa$-deformation.

We begin our discussion by considering a source and a test particle of masses $M$ and $m$, respectively. Now let us assume that the source $M$ (located at the center) is surrounded by a spherically symmetric surface $\mathcal{S}$ (also known as a holographic screen) such that the test particle $m$ lies outside the surface $\mathcal{S}$ and remains very close to $\mathcal{S}$ \cite{verlinde}. 

The energy of this surface $\mathcal{S}$ can be written using the law of equipartition of energy as $E_{\mathcal{S}}=\frac{1}{2}NT$, where $T$ is the equilibrium temperature of the surface $\mathcal{S}$ and $N$ is the number of degree of freedom on the surface $\mathcal{S}$ \cite{verlinde}. This $N$ is related to the surface area ($A$) of $\mathcal{S}$ as $A=QN$, where $Q=\beta l_p^2$ \cite{paddy}. Note that here $l_p$ is the Planck length and $\beta$ is a dimensionless factor which will be fixed later.

In order to define the entropic force, we need an expression for the entropy associated with $\mathcal{S}$. So now we consider a general expression for the entropy ($S$) of the surface $\mathcal{S}$ (in terms of the surface area) as \cite{sheykhi,nicolini,modesto}
\begin{equation}\label{b2}
  S=\frac{A}{4G}+s(A).
\end{equation}
In the above expression, $s(A)$ represents the modification to the entropy which depends on the surface area $A$ and this term is expected to contain the quantum gravity corrections. As a result $s(A)$ becomes significant around the Planck energy scale and therefore this term $s(A)$ vanishes when we approach the low energy limit. In this work $s(A)$ plays a crucial role as this is related to the length scale of the $\kappa$-Minkowski space-time.

According to \cite{verlinde}, the Newton's force acting between the source and the test particle is caused due to the change in the entropy of $\mathcal{S}$ with respect to the displacement of this test particle (i.e., $\Delta x$) from $\mathcal{S}$, which is at an equilibrium temperature $T$. This force expression is given as \cite{verlinde}
\begin{equation}\label{b3}
 F=T\frac{\Delta S}{\Delta x}.
\end{equation}
Substituting Eq.(\ref{b2}) in Eq.(\ref{b3}), we get the expression for force as
\begin{equation}\label{b4}
 F=T\Big(\frac{1}{4G}+\frac{\partial s}{\partial A}\Big)\frac{\Delta A}{\Delta x}
\end{equation}
From $A=QN$, we find $\Delta A=Q\Delta N$ and by setting $\Delta N\sim 1$, we get $\Delta A\sim Q$ \cite{sheykhi}. Now we consider the equilibrium temperature $T$ as the Unruh temperature, i.e., $T=\frac{\mathcal{A}}{2\pi}$ (where $\mathcal{A}$ is the proper acceleration) and the displacement of the test particle from $\mathcal{S}$ as the Compton wavelength, i.e., $\Delta x\sim\lambda_m$ (where $\lambda_m=\frac{1}{m}$). By using these relations in Eq.(\ref{b4}) and by setting $\beta=8\pi$, we get the entropic force as
\begin{equation}\label{b5}
 F=m\mathcal{A}\Big(1+4G\frac{\partial s}{\partial A}\Big).
\end{equation} 
The above equation gives the expression for the Newton's law with an entropic correction term. Note that when we go down the energy scale this entropic correction term $s(A)$ vanishes and we get the standard expression for the Newton's law as $F=m\mathcal{A}$.

Here we are interested in finding the explicit form of $s(A)$ caused due to the $\kappa$-deformation of the space-time. In order to arrive this we take an alternate way by considering the expression of $\kappa$-deformed Unruh temperature instead of considering the expression for corrected entropy. By substituting the expression of $\kappa$-deformed Unruh temperature as $T=\frac{\mathcal{A}}{2\pi}\big(1+\frac{\alpha\mathcal{A}}{4\pi}\big)$ (valid up to first order in the deformation parameter $\alpha$) \cite{vr1} and using $S=\frac{A}{4G}$ (instead of the Eq.(\ref{b2})) along with the previous relations (i.e., $\Delta N\sim 1$, $\Delta x\sim\lambda_m$, $\beta=8\pi$) in Eq.(\ref{b3}), we get the equation for the entropic force in $\kappa$-deformed space-time as
\begin{equation}\label{b6}
 F=m\mathcal{A}\Big(1+\frac{\alpha\mathcal{A}}{4\pi}\Big).
\end{equation}
Eq.(\ref{b6}) is the expression for the $\kappa$-deformed Newton's law. Here this correction term appears due to the $\kappa$-deformation of the Unruh temperature. As in the previous case Eq.(\ref{b6}) will also reduce to the standard form of Newton's equation in the limit $\alpha\to 0$. 

We can now consider this surface $\mathcal{S}$ to be an equipotential surface endowed with a fixed Newton's potential. As a result the test particle close to the surface $\mathcal{S}$ will have a local acceleration due to Newton's gravity. Thus we identify $\mathcal{A}$ as the acceleration due to the Newton's gravity, i.e., $\mathcal{A}=-\frac{GM}{r^2}$. By substituting this definition in Eq.(\ref{b6}), we obtain the expression for the Newton's gravitational force in $\kappa$-deformed space-time as
\begin{equation}\label{b8}
 F=-\frac{GMm}{r^2}\Big(1-\frac{\alpha GM}{4\pi r^2}\Big).
\end{equation} 
From Eq.(\ref{b8}) we obtain the $\kappa$-corrected Newton's potential as
\begin{equation}\label{b9}
 V=-\frac{GM}{r}\Big(1-\frac{\alpha GM}{12\pi r^2}\Big).
\end{equation}
The first order leading correction term to the Newton's potential depends on the square of the mass of the gravitational source and on the radial distance as $1/r^3$. Due to the $1/r^3$ dependency, the effect of minimal length scale to will be dominant in small distance limit. This behaviour is expected as one can observe the effects of non-commutativity around the Planck scale. 

We obtain an explicit form for the correction term $s$, by comparing Eq.(\ref{b5}) with Eq.(\ref{b6}), as $s=\frac{\alpha}{2G}\int dr~r\mathcal{A}$. This gives a direct relation between the entropic correction term $s(A)$ and the length scale of the $\kappa$-deformed space-time such that this correction term will vanish when we take the corresponding commutative limit. Substituting $\mathcal{A}=-\frac{GM}{r^2}$ and performing the integral, we obtain an explicit form for $s(A)$. Now using this $s(A)$ in Eq.(\ref{b2}), we get the $\kappa$-corrected entropy as
\begin{equation}\label{b10}
 S=\frac{A}{4G}-\frac{\alpha M}{4}\ln\Big(\frac{A}{4\pi}\Big).
\end{equation}
Thus we obtain a logarithmic correction to the Bekenstein-Hawking entropy as the first order correction term associated with the $\kappa$-deformation of space-time. Moreover this correction also depends on the mass of black hole.

%%%%%%%%%%%%%%%%%%%%%%%%%%%%%%%%%%%%%%%%%%%%%%%%%%%%%%%%%%%%%%%%%%%%%%%%%%%%%%%%%%%%%%%%%%%%%%%%%%%%%%%%%%%%%%%%%%%%%%%%%%%%%%%%%%%%%%%%%%%%%%%%%%%%

\section{$\kappa$-modified Friedmann's equation and bouncing scenario}

The Friedmann equations provide the dynamical equations for studying the evolution of early universe. Interestingly this Friedmann equations can be obtained directly from the Newton's gravitational force expression. This can be achieved by considering the radius $r$ of the holographic screen $\mathcal{S}$ in terms of the scale factor $a$ of universe, i.e., $r(t)=a(t)r_0$ (where $r_0$ is the present radius of universe) and by replacing the source mass $M$ in the force expression with the corresponding gravitational mass.  So in this section we derive the Friedmann equations in $\kappa$-Minkowski space-time from the $\kappa$-deformed Newton's gravitational force equation obtained in Eq.(\ref{b8}). Using this modified Friedmann equations we then study the bouncing scenario in the $\kappa$-deformed space-time. 

We consider a homogeneous, isotropic universe described by the FLRW metric as
\begin{equation}\label{c1}
 ds^2=h_{\mu\nu}dx^{\mu}dx^{\nu}+r^2d\Omega^2,
\end{equation}
where $h_{\mu\nu}=$diag$(-1,a^2(t)/(1-kr_0^2))$ and $x^{\mu}=(t,r_0)$. Here $k$ represents the curvature of space, where $k=0,1,-1$ correspond to spatially flat, closed, and open universes, respectively. The radius of the dynamical apparent horizon of FLRW universe can be calculated from $h^{\mu\nu}\partial_{\mu}r\partial_{\nu}r=0$ as
\begin{equation}\label{c1a}
 r=\frac{1}{\sqrt{H^2+k/a^2}},
\end{equation}
where $H$ is the Hubble parameter defined as $H=\dot{a}/a$.

The matter content of this FRW metric is described using the energy-momentum tensor of the perfect fluid as
\begin{equation}\label{c2}
 T_{\mu\nu}=\big(\rho+p\big)u_{\mu}u_{\nu}+pg_{\mu\nu},
\end{equation} 
where $\rho$ is the density, $p$ is the pressure and $u_{\mu}$ is the unit $4$-velocity associated with the perfect fluid. The evolution of this perfect fluid is governed by the continuity equation $\dot{\rho}+3H\big(1+\omega\big)\rho=0$, where $\omega$ is the equation of state parameter given by $\omega=p/\rho$. By solving this we get the expression for pressure as $\rho=\rho_0 a^{-3(1+\omega)}$, where $\rho_0$ is the present density of universe. 

Before proceeding to derive the Friedmann equation from Eq.(\ref{b8}), it is important to understand that $M$ in Eq.(\ref{b8}) represents the total mass of the system rather than its gravitational mass. So in order to derive the Friedmann equation, we need to replace $M$ in Eq.(\ref{b8}) with the gravitational mass $\mathcal{M}$. The explicit form of this gravitational mass $\mathcal{M}$ can be calculated using the Komar mass, i.e., $\mathcal{M}=2\int dV~\Big(T_{\mu\nu}-\frac{1}{2}Tg_{\mu\nu}\Big)u^{\mu}u^{\nu}$. Thus by substituting Eq.(\ref{c1}) and Eq.(\ref{c2}) in the definition for the Komar mass, we get the gravitational mass as
\begin{equation}\label{c3}
 \mathcal{M}=\frac{4\pi r^3}{3}\big(\rho+3p\big). 
\end{equation}
By replacing $M$ in Eq.(\ref{b8}) with the above expression for gravitational mass, i.e., Eq.(\ref{c3}), we get the $\kappa$-corrected first Friedmann equation as
\begin{widetext}
\begin{equation}\label{c7}
 \frac{\ddot{a}}{a}=-\frac{4\pi G}{3}\big(1+3\omega\big)\rho_0a^{-3(1+\omega)}\bigg[1-\frac{\alpha G\big(1+3\omega\big)\rho_0a^{-3(1+\omega)+1} r_0}{3}\bigg],
\end{equation} 
\end{widetext}
where we use the relations $p=\omega\rho$, $\rho=\rho_0a^{-3(1+\omega)}$ and $r=a r_0$. Multiplying Eq.(\ref{c7}) throughout by $2a\dot{a}$ and integrating the resulting equation we get the $\kappa$-corrected second Friedmann equation as
\begin{widetext}
\begin{equation}\label{c8}
 H^2+\frac{k}{a^2}=\frac{8\pi G}{3}\rho_0a^{-3(1+\omega)}\bigg[1-\frac{\alpha G}{9}\frac{(1+3\omega)^2}{(1+2\omega)}\rho_0a^{-(2+3\omega)}r_0\bigg],
\end{equation}
\end{widetext}
where $k$ is the integration constant and is interpreted as the spatial curvature. After a small re-arrangement using $\rho=\rho_0a^{-3(1+\omega)}$ and Eq.(\ref{c1a}), we obtain the $\kappa$-corrected Friedman equations as
\begin{equation}\label{c9}
 \frac{\ddot{a}}{a}=-\frac{4\pi G}{3}\big(1+3\omega\big)\rho \Bigg(1-\frac{\alpha G}{3}\big(1+3\omega\big) \frac{\rho}{\sqrt{H^2+\frac{k}{a^2}}} \Bigg),
\end{equation}
\begin{equation}\label{c10}
 H^2+\frac{k}{a^2}=\frac{8\pi G}{3}\rho \Bigg(1-\frac{\alpha G}{9}\frac{(1+3\omega)^2}{(1+2\omega)}\frac{\rho}{\sqrt{H^2+\frac{k}{a^2}}} \Bigg).
\end{equation}
Up to first order term in $\alpha$, $\sqrt{H^2+\frac{k}{a^2}}$ term in the RHS of Eq.(\ref{c9}) and Eq.(\ref{c10}) can be re-cast using $\rho$ as
\begin{equation}\label{c11}
 \frac{\ddot{a}}{a}=-\frac{4\pi G}{3}\big(1+3\omega\big)\rho \bigg(1-\frac{\alpha G}{3} \sqrt{\frac{3}{8\pi G}} \big(1+3\omega\big) \sqrt{\rho} \bigg),
\end{equation}
\begin{equation}\label{c12}
 H^2+\frac{k}{a^2}=\frac{8\pi G}{3}\rho \bigg(1-\frac{\alpha G}{9} \sqrt{\frac{3}{8\pi G}} \frac{(1+3\omega)^2}{(1+2\omega)} \sqrt{\rho} \bigg).
\end{equation}
In both Eq.(\ref{c11}) and Eq.(\ref{c12}), we observe a non-trivial $\rho^{3/2}$ dependent correction term due to the space-time non-commutativity. Such higher power density dependent correction terms have been shown to appear in the Friedmann equations having length scale corrections \cite{snyder-bounce,moyal-bounce,gup-bounce,alonso,jusufi,yi-ling,mdr-bounce,rainbow-bounce}. All these modified Friedmann equations reduce to the standard ones in the vanishing limit of the corresponding length scales.

%%%%%%%%%%%%%%%%%%%%%%%%%%%%%%%%%%%%%%%%%%%%%%%%%%%%%%%%%%%%%%%%%%%%%%%%%%%%%%%%%%%%%%%%%%%%%%%%%%%%%%%%%%%%%%%%%%%%%%%%%%%%%%%%%%%%%%%%%%%

\subsection{Evolution of scale factor in $\kappa$-deformed space-time}

This correction term plays a crucial role in the evolution of universe at early times. The effect of the non-commutativity in early universe can be analysed by studying the evolution of scale factor carrying non-commutative correction terms. For this we first obtain the expression for Hubble parameter in $\kappa$-Minkowski space-time from the following equation
\begin{equation}\label{c13}
 \dot{H}=-H^2+\frac{\ddot{a}}{a}.
\end{equation}
By substituting Eq.(\ref{c9}) in Eq.(\ref{c13}) and re-writting the $\rho$ using Eq.(\ref{c10}), we get 
\begin{widetext}
\begin{equation}\label{c14}
 \dot{H}=-H^2-\frac{1}{2}(1+3\omega) \Big(H^2+\frac{k}{a^2}\Big) \bigg(1-\frac{a(1+3\omega)(2+3\omega)}{24\pi(1+2\omega)}\sqrt{H^2+\frac{k}{a^2}}\bigg)
\end{equation}
\end{widetext}
Now we consider the case of spatially flat universe $k=0$ and hence Eq.(\ref{c14}) becomes
\begin{equation}\label{c15}
 \dot{H}=-\frac{3}{2}H^2(1+\omega) +\frac{\alpha(1+3\omega)^2(2+3\omega)}{48\pi(1+2\omega)}H^3.
\end{equation}
We look for the first order perturbative solution by expanding $H(t)$ valid up to $\alpha$ as
\begin{equation}\label{c16}
 H(t)=H^{(0)}(t)+\alpha H^{(\alpha)}(t).
\end{equation}
$H^{(0)}$ is the standard solution when non-commutative effects are absent and is obtained to be $H^{(0)}(t)=\frac{2}{3(1+\omega)t}$. On the other hand $H^{(\alpha)}$ is the first order correction that appears when we consider the non-commutative effects. By using Eq.(\ref{c16}) and $H^{(0)}(t)=\frac{2}{3(1+\omega)t}$ in Eq.(\ref{c15}), we obtain a differential equation for $H^{(\alpha)}(t)$, as $\frac{d{H}^{(\alpha)}(t)}{dt}+\frac{2{H}^{(\alpha)}(t)}{t}=\frac{4(2+3\omega)(1+3\omega)^2}{81\pi(1+\omega)^3(1+2\omega)t^3}$.
After obtaining the solution for this differential equation and substituting it back in Eq.(\ref{c16}), we get the expression for the Hubble parameter in $\kappa$-deformed space-time as
\begin{equation}\label{c18}
 H(t)=\frac{2}{3(1+\omega)t}+\frac{4\alpha(1+3\omega)^2(2+3\omega)}{81\pi(1+2\omega)}\frac{\ln t}{t}.
\end{equation}
Using $H=\frac{\dot{a}}{a}$ and integrating the Eq.(\ref{c18}), we get the scale factor in $\kappa$-Minkowski space-time as
\begin{widetext}
\begin{equation}\label{c19}
 {a(\tau)}=\tau^{2/3(1+\omega)}\bigg(1-\frac{4\alpha(1+3\omega)^2(2+3\omega)}{81\pi(1+2\omega)}\frac{(1+\ln \tau)}{\tau}\bigg),
\end{equation}
\end{widetext}
where we denote $\tau={t}/{t_0}$ as the normalised time.

\begin{figure}[!htb]\centering
\includegraphics[width=0.50\textwidth]{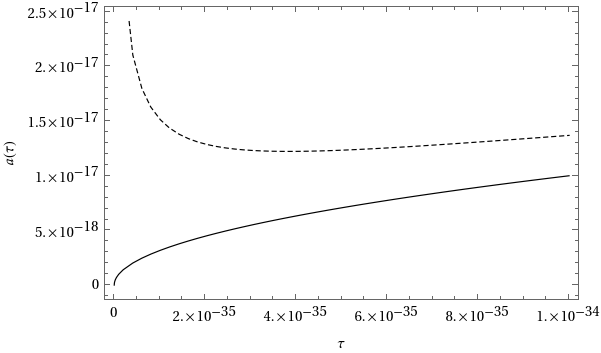}
\caption{Plot of the scale factor in a radiation dominated ($\omega=1/3$) universe against $\tau$. The dashed line depicts the $\kappa$-corrected scale factor with $\alpha=10^{-35}m$ and solid line corresponds to the standard scale factor with $\alpha=0$.} 
\label{fig:rad1}
\end{figure}

For a radiation dominated epoch $\omega=1/3$, the $\kappa$-corrected scale factor takes the form
\begin{equation}\label{c20}
 {a(\tau)}=\tau^{1/2}-\frac{3\alpha}{20\pi}\frac{1}{\tau^{1/2}}-\frac{3\alpha}{20\pi}\frac{\ln \tau}{\tau^{1/2}}
\end{equation}
The first order correction contains $\tau^{-1/2}$ as well as $\tau^{-1/2}\ln \tau$ dependent terms and as a result the scale factor diverges as we approach $\tau\to 0$. Also this scale factor asymptotically attains a finite value as $\tau\to\infty$. In Fig.(\ref{fig:rad1}) we find these correction terms avoid the initial singularity by inducing a bounce behaviour. Similar bounce behaviour is shown to appear for radiation dominated era in Moyal space-time \cite{moyal-bounce}. 

\begin{figure}[!htb]\centering
\includegraphics[width=0.50\textwidth]{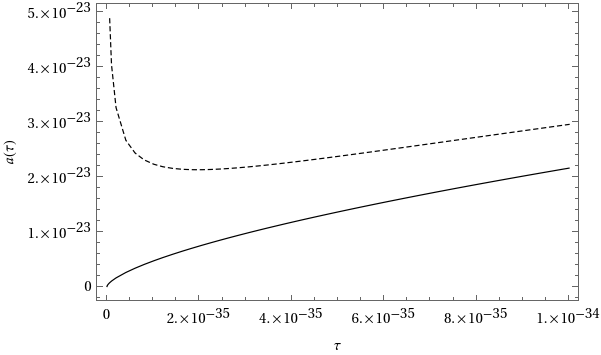}
\caption{Plot of the scale factor in a matter dominated ($\omega=0$) universe against $\tau$. The dashed line depicts the $\kappa$-corrected scale factor with $\alpha=10^{-35}m$ and solid line corresponds to the standard scale factor with $\alpha=0$.} 
\label{fig:matter1}
\end{figure}

For a matter dominated era $\omega=0$, the scale factor becomes
\begin{equation}\label{c21}
 {a(\tau)}=\tau^{2/3}-\frac{8\alpha}{81\pi}\frac{1}{\tau^{1/3}}-\frac{8\alpha}{81\pi}\frac{\ln \tau}{\tau^{1/3}}.
\end{equation}
Here the first order correction terms depend on $\tau^{-1/3}$ as well as $\tau^{-1/3}\ln \tau$ and these terms make the scale factor to diverge in the limit $\tau\to 0$. As in the previous case here also the scale factor asymptotically attains a finite value in the limit $\tau\to\infty$. From Fig.(\ref{fig:matter1}) we find that the $\kappa$-deformed correction terms removes the initial singularity by providing a bouncing solution.

%%%%%%%%%%%%%%%%%%%%%%%%%%%%%%%%%%%%%%%%%%%%%%%%%%%%%%%%%%%%%%%%%%%%%%%%%%%%%%%%%%%%%%%%%%%%%%%%%%%%%%%%%%%%%%%%%

\subsection{Bouncing scenario in $\kappa$-deformed space-time}

Thus from Fig.(\ref{fig:rad1}) and Fig.(\ref{fig:matter1}), the minimal length scale associated with the $\kappa$-deformed space-time is seen to avoid the initial singularity in the early universe by providing a bouncing behaviour. Now we check whether scalar factor and density satisfy the bouncing conditions or not. We analyse this using the modified Friedmann equations derived in Eq.(\ref{c11}) and Eq.(\ref{c12}).

Due to the bouncing behaviour the universe initially undergoes contraction (where $\dot{a}<0$) followed by the expansion (where $\dot{a}>0$), unlike the standard big bang model where the universe begins with an initial singularity. As the universe contracts the scale factor reduces to a non-zero minimum value, after which it bounces back and starts to expand again. This minimal scale factor is called the critical point. We denote the scale factor at critical point as $a_b$. At this critical point, the scale factor satisfy $\dot{a}_b=0$ along with the conditions $a_b>0$ and $\ddot{a}_b>0$. 

Thus at critical point we have $H\big|_{a_b}=\frac{\dot{a}_b}{a_b}=0$ and hence the $\kappa$-deformed Friedmann equation given in Eq.(\ref{c12}) becomes
\begin{equation}\label{e1}
 \frac{k}{a_b^2}=\frac{8\pi G}{3}\rho_b\bigg(1-\frac{\alpha G}{9}\sqrt{\frac{3}{8\pi G}}\frac{(1+3\omega)^2}{(1+2\omega)}\sqrt{\rho_b}\bigg)
\end{equation}
where $\rho_b$ is the density at the critical point. By taking the square root of both side and expanding the RHS by keeping the first order terms in $\alpha$, Eq.(\ref{e1}) becomes
\begin{equation}\label{e1a}
 \frac{\sqrt{k}}{a_b}=\sqrt{\frac{8\pi G}{3}}\sqrt{\rho_b}\bigg(1-\frac{\sqrt{\rho_b}}{\sqrt{\rho_{\kappa}}}\bigg),
\end{equation}
where we denote $\sqrt{\rho_{\kappa}}=\frac{18}{\alpha G}\frac{(1+2\omega)}{(1+3\omega)^2}\sqrt{\frac{8\pi G}{3}}$ such that $\rho_{\kappa}$ diverges in the commutative limit $\alpha\to 0$. The above expression is seen to be quadratic equation in $\sqrt{\rho}_b$ and its solutions can be written as
\begin{equation}\label{e2}
 \sqrt{\rho_b}=\frac{1}{2} \sqrt{\rho_{\kappa}} \bigg( 1 \pm \sqrt{1-\frac{\alpha\sqrt{k}}{12\pi a_b}\frac{(1+3\omega)^2}{(1+2\omega)}}  \bigg).
\end{equation} 
Using $\sqrt{\rho_{\kappa}}$ in Eq.(\ref{c11}), the modified Friedmann equation at bounce will be
\begin{equation}\label{e3}
 \frac{\ddot{a}_b}{a_b}=-\frac{4\pi G}{3}\big(1+3\omega\big)\rho_b \bigg( 1 - \frac{6(1+2\omega)}{(1+3\omega)}\frac{\sqrt{\rho_b}}{\sqrt{\rho_{\kappa}}}   \bigg).
\end{equation}  
Now we analyse the bouncing conditions from Eq.(\ref{e3}) for different values of $k$. So first let us consider the case of $k=0$ at which we obtain two solutions for ${\rho}_b$ from Eq.(\ref{e2}) as ${\rho_b^{(+)}}={\rho_{\kappa}}$ and ${\rho_b^{(-)}}=0$, respectively. Substituting $\rho_b^{(+)}=\rho_{\kappa}$ in Eq.(\ref{e3}), we find
\begin{equation}\label{e4}
 \frac{\ddot{a}_b}{a_b}=\frac{1152\pi^2}{\alpha^2}\frac{(1+2\omega)^2(5+9\omega)}{(1+3\omega)^4}.
\end{equation}
Here $\frac{\ddot{a}_b}{a_b}>0$ provided $\omega>-\frac{5}{9}$. But on the other hand $\frac{\ddot{a}_b}{a_b}$ vanishes for $\rho_b^{(-)}$. Thus we observe a bounce behaviour only for $\rho_b^{(+)}$ in $k=0$, as seen in Fig.(\ref{fig:rad1}) and Fig.(\ref{fig:matter1}) (for radiation and matter dominated eras). Further it is obvious from Eq.(\ref{e2}) that one cannot have a bounce for $k=-1$ as $\rho_b$ becomes an imaginary number.

For $k=1$ in Eq.(\ref{e2}), we get the solutions for ${\rho}_b$ as $\rho_b^{(+)}=\rho_{\kappa}\Big(1-\frac{\alpha}{24\pi a_b}\frac{(1+3\omega)^2}{(1+2\omega)}\Big)$ and $\rho_b^{(-)}=\frac{3}{8\pi G a_b^2}$. By using $\rho_b^{(+)}$ in Eq.(\ref{e3}), we get 
\begin{widetext}
\begin{equation}\label{e5}
\frac{\ddot{a}_b}{a_b}=\frac{48\pi}{\alpha^2a_b}  \frac{\big(24\pi a_b(5+9\omega)(1+2\omega)-\alpha(1+3\omega)^2(8+15\omega)\big)(1+2\omega)}{(1+3\omega)^4}.
\end{equation}
\end{widetext}
Here we find $\frac{\ddot{a}_b}{a_b}>0$ only when $a_b$ and $\omega$ follow the conditions $a_b > \frac{\alpha(1+3\omega)^2(8+15\omega)}{24\pi(5+9\omega)(1+2\omega)}$ and $\omega>-\frac{1}{2}$, respectively. Similarly by using $\rho_b^{(-)}$ in Eq.(\ref{e3}), we get 
\begin{equation}\label{e6}
\frac{\ddot{a}_b}{a_b}=\frac{(1+3\omega)\big(\alpha(1+3\omega)-8\pi a_b\big)}{16\pi a_b^3}.
\end{equation}
Here $\frac{\ddot{a}_b}{a_b}>0$ when $a_b < \frac{\alpha(1+3\omega)}{8\pi}$ provided $\omega>-\frac{1}{3}$. Thus one can observe a bounce in $k=1$ when the conditions $\frac{\alpha(1+3\omega)^2(8+15\omega)}{24\pi(5+9\omega)(1+2\omega)} < a_b < \frac{\alpha(1+3\omega)}{8\pi}$ and $\omega>-\frac{1}{3}$ are satisfied. So in order to observe the bounce in a radiation dominated epoch the $a_b$ should be in the interval $(\frac{13\alpha}{80\pi},\frac{2\alpha}{8\pi})$. Similarly one can also observe the bounce in a matter dominated era when $a_b$ comes within the range $(\frac{\alpha}{15\pi},\frac{\alpha}{8\pi})$. From these analysis it is evident that the bouncing behaviour is completely contributed by the minimal length scale associated with the $\kappa$-Minkowski space-time. Thus in the commutative case (i.e., $\lim \alpha\to 0$) one cannot have a bounce resulting in an initial singularity as seen in the standard cosmology.

%%%%%%%%%%%%%%%%%%%%%%%%%%%%%%%%%%%%%%%%%%%%%%%%%%%%%%%%%%%%%%%%%%%%%%%%%%%%%%%%%%%%%%%%%%%%%%%%%%%%%%%%%%%%%%%%%
\section{Conclusion}

We have generalised the concept of entropic force into the $\kappa$-deformed space-time and have further used it to derive the expression for Newton's force of gravity in the $\kappa$-Minkowski space-time by considering the expression for the $\kappa$-corrected Unruh temperature, obtained in \cite{vr1}, as the equilibrium temperature of the holographic screen. The first order corrections to the Newton's gravitational force in $\kappa$-deformed space-time is seen to be a $1/r^4$ dependent term. Our expression for the $\kappa$-corrected Newton's potential is similar to the results obtained in \cite{qnc} as these expressions for the quantum corrected Newton's potential also contains a $1/r^3$ term. The departure from standard inverse square potential becomes significant at small distance scales signals the remnants of quantum gravity effects, which arise from the modifications to space-time geometry. 

In this study we have shown the appearance of logarithmic correction as the first order $\kappa$-deformed correction term of the Bekenstein-Hawking entropy. Here our results are valid even for the uncharged black holes in comparison with the one derived in \cite{tajron}, where the logarithmic corrections due to non-commutativity has been shown to appear only for the charged black holes. However in both these cases the logarithmic correction term appears as the non-commutative correction associated with the entropy, in contrast with the commutative case where such logarithmic correction occurs as the quantum corrections to the entropy. It is worth mentioning that similar logarithmic corrections to the black hole entropy has also been obtained in various quantum gravity approaches \cite{sen,xiao}. Therefore one can possibly expect the NC space-time structures to encapsulate the quantum gravity features inherently. 

We have derived the Friedmann equations in $\kappa$-deformed space-time from the expression for the $\kappa$ corrected Newton's gravitational force and the leading correction term is found to have a non-trivial dependence on the density, in contrast to the linear power density dependent correction term obtained in \cite{suman}. Similar non-trivial or higher power density dependent correction terms have been shown to play a crucial role in the early universe cosmology.

Using this modified Friedmann equations we have studied the evolution of the $\kappa$ corrected scale factor in a spatially flat universe and find that the $\kappa$-deformation parameter avoids the initial singularity by introducing a bouncing behaviour. This bouncing behaviour is in close agreement with the one observed in Moyal space-time \cite{moyal-bounce}. The non-singular bouncing behaviour has also been shown to appear in the context of Snyder NC space-time and interestingly this Snyder deformed Heisenberg algebra can be related to the $\kappa$-Poincare algebra \cite{snyder-bounce}. The deformed Friedman equations in LQG framework is shown to provide a non-singular bouncing universe \cite{bojowald,abhay}. Moreover in the low energy limit of certain loop quantum gravity, symmetry algebra of the resulting space-time can be described using $\kappa$-Poincare algebra \cite{loop-kappa} and thus it is quite natural to expect a quantum bounce in $\kappa$-Minkowski space-time as well.

From our further analysis we have shown that the bouncing behaviour due to the $\kappa$-deformation is possible only in the spatially flat and closed universes. We find that the bounce in $k=0$ occurs for $\omega>-\frac{5}{9}$ and that in $k=1$ happens when the scale factor at the critical point obeys $\frac{\alpha(1+3\omega)^2(8+15\omega)}{24\pi(5+9\omega)(1+2\omega)} < a_b < \frac{\alpha(1+3\omega)}{8\pi}$ for $\omega>-\frac{1}{3}$. Our result is similar to the one observed in \cite{mdr-bounce}, where the bounce appears only for $k=0$ and $k=1$, due to the modified dispersion relation based on DSR. Moreover the $\kappa$-Minkowski space-time happens to be the underlying the space-time for the DSR theories. Thus our result is in consistent with the above mentioned ones.

Thus it is intuitive to think that quantum space-time structures endowed with a minimal length scale could possibly avoid the initial singularity in the early universe by providing a bouncing behaviour or some other alternate mechanism. But still we need a complete picture of the quantum gravity theory to make a conclusive statement regarding this.  

\section*{Acknowledgment}

This work is supported by the National Natural Science Foundation of China (Grant No.~12275080), and the Innovative Research Group of Hunan Province (Grant No.~2024JJ1006).

%%%%%%%%%%%%%%%%%%%%%%%%%%%%%%%%%%%%%%%%%%%%%%%%%%%%%%%%%%%%%%%%%%%%%%%%%%%%%%%%%%%%%%%%%%%%%%%%%%%%%%%%%%%%%%%%%

%\section*{}


\begin{thebibliography}{99}

\bibitem{ml} S. Hossenfelder, \textit{Living Rev. Relativ. }\textbf{16} (2013) 2.

\bibitem{bojowald} M. Bojowald, \textit{Phys. Rev. Lett. }\textbf{86} (2001) 5227.

\bibitem{abhay} A. Ashtekar, T. Pawlowski and P. Singh, \textit{Phys. Rev. Lett. }\textbf{96} (2006) 141301.

\bibitem{connes} A. Connes, \textit{Noncommutative geometry}, Academic Press (1994).

\bibitem{doplicher} S. Doplicher, K. Fredenhagen and J. E. Roberts, \textit{Phys. Lett. }\textbf{B 331} (1994) 39; S. Doplicher, K. Fredenhagen and J. E. Roberts, \textit{Comm. Math. Phys. }\textbf{172} (1995) 187.

\bibitem{kappa-review} A. Borowiec and A. Pachol, \textit{SIGMA} \textbf{6} (2010) 086; M. Arzano and J. K.-Glikman, \textit{Symmetry} \textbf{13} (2021) 946;

\bibitem{kpa} J. Lukierski, A. Nowicki, H. Ruegg, and V. N. Tolstoy, \textit{Phys. Lett. }\textbf{B 264} (1991) 331; J. Lukierski, A. Nowicki, and H. Ruegg, \textit{Phys. Lett. }\textbf{B 293} (1992) 344; J. Lukierski and H. Ruegg, \textit{Phys. Lett. }\textbf{B 329} (1994) 189; S. Majid and H. Ruegg, \textit{Phys. Lett. }\textbf{B 334} (1994) 348.

\bibitem{dsr} G. A.-Camelia, \textit{Special treatment. Nature} \textbf{418} (2002) 34; J. K.-Glikman, \textit{Lect. Notes Phys. }\textbf{669} (2005) 131.

\bibitem{meljanac} S. Meljanac and M. Stojic, \textit{Eur. Phys. J. }\textbf{C 47} (2006) 531; S. K.-Juric, S. Meljanac, and M. Stojic, \textit{Eur. Phys. J. }\textbf{C 51} (2007) 229.

\bibitem{kumar} S. Meljanac, A. Samsarov, M. Stojic and K. S. Gupta, \textit{Eur. Phys. J. }\textbf{C 53} (2008) 295.

\bibitem{kappa-gravity} E. J. Beggs and S. Majid, \textit{Class. Quant. Grav. }\textbf{31} (2014) 035020.

\bibitem{kappa-gravity1} N. Herceg, T. Juric, A. Samsarov and I. Smolic, \textit{JHEP} \textbf{06} (2024) 130.

\bibitem{ehk} E. Harikumar and A. K. Kapoor, \textit{Mod. Phys. Lett. }\textbf{A 25} (2010) 2991.

\bibitem{ehk1} E. Harikumar, T. Juric and S. Meljanac, \textit{Phys. Rev. }\textbf{D 84} (2011); E. Harikumar and V. Rajagopal, \textit{Ann. Phys. }\textbf{423} (2020) 168332; E. Harikumar, L. G. C. Lakkaraju and V. Rajagopal, \textit{Mod. Phys. Lett. }\textbf{A 36} (2021) 2150069.

\bibitem{suman} E. Harikumar, H. Sreekumar and S. K. Panja, \textit{Universe} \textbf{9} (2023) 343.

\bibitem{snyder-bounce} M. V. Battisti, \textit{Phys. Rev. }\textbf{D 79} (2009) 083506.

\bibitem{moyal-bounce} R. Matsuyama and M. Nagasawa, \textit{Class. Quant. Grav. }\textbf{35} (2018) 155010.

\bibitem{gup-bounce} M. Salah, F. Hammad, M. Faizal and A. F. Ali, \textit{JCAP} \textbf{02} (2017) 035.

\bibitem{alonso} A. A.-Serrano, M. Liska and A. V.-Becerril, \textit{Phys. Lett. }\textbf{B 839} (2023) 137827.

\bibitem{yi-ling} Y. Ling, W. J. Li and J.-P. Wu, \textit{JCAP} \textbf{11} (2009) 016.

\bibitem{rainbow-bounce} Y. Ling and Q. Wu, \textit{Phys. Lett. }\textbf{B 687} (2010) 103.

\bibitem{mdr-bounce} W.-J. Pan and Y.-C. Huang, \textit{Gen. Rel. Grav. }\textbf{48} (2016) 144.

\bibitem{jusufi} K. Jusufi and A. Sheykhi, \textit{Phys. Lett. }\textbf{B 836} (2023) 137621.

\bibitem{verlinde} E. Verlinde, \textit{JHEP} \textbf{04} (2011) 029.

\bibitem{gup-ef} S. Kibaroglu and M. Senay, \textit{Int. J. Mod. Phys. }\textbf{D 29} (2020) 2050042.

\bibitem{e-f} R.-G. Cai, L.-M. Cao and N. Ohta, \textit{Phys. Rev. }\textbf{D 81} (2010) 061501.

\bibitem{sheykhi} A. Sheykhi, \textit{Phys. Rev. }\textbf{D 81} (2010) 104011.

\bibitem{vr1} V. Rajagopal, \textit{Gen. Rel. Grav. }\textbf{56} (2024) 14.

\bibitem{paddy} T. Padmanabhan, \textit{Mod. Phys. Lett. }\textbf{A 25} (2010) 1129; \textit{Phys. Rev. }\textbf{D 81} (2010) 124040.

\bibitem{nicolini} P. Nicolini, \textit{Phys. Rev. }\textbf{D 82} (2010) 044030.

\bibitem{modesto} L. Modesto and A. Randono, arXiv:1003.1998 [hep-th].

\bibitem{qnc} M. J. Duff, \textit{Phys. Rev. }\textbf{D 9} (1974) 1837; J. F. Donoghue, \textit{Phys. Rev. }\textbf{D 50} (1994) 3874; H. W. Hamber and S. Liu, \textit{Phys. Lett. }\textbf{B 357} (1995) 51; I. J. Muzinich and S. Vokos, \textit{Phys. Rev. }\textbf{D 52} (1995) 3472; I. B. Khriplovich and G. G. Kirilin, \textit{J. Exp. Theor. Phys. }\textbf{95} (2002) 981; N. E. J. B.-Borh, J. F. Donoghue and B. R. Holstein, \textit{Phys. Rev. }\textbf{D 67} (2003) 084033;  N. E. J. B.-Borh, J. F. Donoghue and B. R. Holstein, \textit{Phys. Rev. }\textbf{D 68} (2003) 084005; G. G. Kirilin, \textit{Phys. Rev. }\textbf{D 75} (2007) 108501; N. E. J. B.-Bohr, J.F. Donoghue, B. R. Holstein, L. Plante and P. Vanhove, \textit{Phys. Rev. Lett. }\textbf{114} (2015) 061301; C. L. Wang and R. P. Woodard, \textit{Phys. Rev. }\textbf{D 92} (2015) 084008.

\bibitem{tajron} A. Hrelja, T. Juric and F. Pozar, arXiv:2407.13233 [hep-th]; T. Juric and F. Pozar, \textit{Symmetry} \textbf{15(2)} (2023) 417; K. S. Gupta, T. Juric, A. Samsarov and I. Smolic, \textit{JHEP} \textbf{02} (2023) 060.

\bibitem{sen} A. Sen, \textit{Gen. Rel. Grav. }\textbf{44} (2012) 1207; \textit{JHEP} \textbf{04} (2013) 156.
\bibitem{xiao} X. Calmet and F. Kuipers, \textit{Phys. Rev. }\textbf{D 104} (2021) 066012; Y. Xiao and Y. Tian, \textit{Phys. Rev. }\textbf{D 105} (2022) 044013. 




\bibitem{loop-kappa} F. Cianfrani, J. K.-Glikman, D. Pranzetti and G. Rosati, \textit{Phys. Rev. }\textbf{D 94} (2016) 084044.








\end{thebibliography}
\end{document}